\documentclass[fleqn,usenatbib]{mnras}

\usepackage{newtxtext,newtxmath}

\usepackage[T1]{fontenc}

\DeclareRobustCommand{\VAN}[3]{#2}
\let\VANthebibliography\thebibliography
\def\thebibliography{\DeclareRobustCommand{\VAN}[3]{##3}\VANthebibliography}

\usepackage{graphicx}   
\usepackage{amsmath}    

\newcommand{\sdo}{{\it SDO}} %
\newcommand{\hinode}{{\it Hinode}} %
\newcommand{\kms}{km\,s$^{-1}$} %
\newcommand{\mac}{magnetoacoustic } %

\title[Dispersiveness of Slow Waves]{Observed Dispersive Properties of the Slow Magnetoacoustic Waves Propagating in Coronal Fan Loops above Sunspots}
\author[Zhao et al.]{
Junwei Zhao,$^{1}$\thanks{E-mail: junwei@sun.stanford.edu}
Tongjiang Wang, $^{2}$
and Ruizhu Chen$^{1}$ \\
$^{1}$ W.~W.~Hansen Experimental Physics Laboratory, Stanford University, Stanford, CA 94305-4085, USA \\
$^{2}$ The Catholic University of America at NASA Goddard Space Flight Center, Greenbelt, MD 20771, USA}

\date{Accepted XXX. Received YYY; in original form ZZZ}
\pubyear{2024}

\begin{document}
\label{firstpage}
\pagerange{\pageref{firstpage}--\pageref{lastpage}}
\maketitle

\begin{abstract}
Recurrent and propagating intensity perturbations are frequently observed in extreme ultraviolet (EUV) channels along coronal fan loops above sunspots, and these perturbations are suggested to be slow magnetoacoustic waves.
Numerous studies have been conducted to investigate their propagation speeds, damping, and excitation sources; however, there have been limited observational analyses on whether these waves are dispersive despite some theoretical studies.
In this study, we apply cross-correlation analysis in the Fourier domain on slow magnetoacoustic waves using three different datasets: EUV intensity observed by \sdo/AIA, differential emission measure (DEM) temperature maps, and Doppler velocities from \hinode/EIS spectrometer observations.
The apparent phase velocities of the waves, which are the plane-of-sky component of the waves' phase velocities, are derived as functions of frequency for all the three datasets.
It is found that the phase velocities show clear frequency dependency, with a general trend of increase with frequency, ranging from approximately 30\,\kms\ around 3\,mHz to about 80\,\kms\ around 10\,mHz.
The frequency dependency of the phase velocities demonstrates that the slow \mac waves in the coronal loops are dispersive.
The dispersiveness of these waves can provide a useful tool for the diagnosis of physical conditions inside the coronal loops along which these waves travel.
\end{abstract}


\section{Introduction}
\label{sec1}

Different types of waves are detected and investigated in and above sunspots, and among them, slow magnetoacoustic waves are actively studied \citep[see recent reviews by][]{wan16, ban21, wan21}.
Apparent propagating perturbations of brightness \citep{sch99, dem02} are observed in the extreme ultraviolet (EUV) coronal fan loops at the temperature around 1.0\,MK by space-based instruments, such as {\it Transition Region And Coronal Explorer} \citep[{\it TRACE};][]{han99} and {\it Solar Dynamics Observatory} / Atmospheric Imaging Assembly \citep[\sdo/AIA;][]{lem12}.
The phenomenon is interpreted by many authors as slow magnetoacoustic waves \citep[e.g.,][]{wan09}.
Although similar phenomena observed in non-spot regions were interpreted by some authors as quasi-periodic upflows \citep[e.g.,][]{dep10}, here we follow the review by \citet{ban21} and discuss the phenomenon associated with sunspots as slow magnetoacoustic waves.

New evidence is accumulating that slow magnetoacoustic waves in coronal fan loops originate from the Sun's photosphere.
Through analyzing the periodicity and propagation speeds of the waves in coronal fans and sunspots, \citet{jes12} found that the slow magnetoacoustic waves were from the lower atmosphere anchoring in the photospheric umbral dots.
By comparing the amplitude modulation periods in different atmospheric layers, \citet{kri15} found that the slow magnetoacoustic waves in the coronal fans were driven by photospheric $p$-modes, which propagate upward into the corona before being dissipated.
Applying a time--distance helioseismic analysis method on multi-wavelength observations, \citet{zha16} traced how the photospheric $p$-mode waves of different frequencies channeled upward through the chromosphere and transition region into the coronal fan loops.
How and where these photospheric waves are generated were also studied by various authors.
\citet{cha17} found that the upwardly propagating slow waves were consistent with the 3-min oscillations observed in light bridges and umbral dots inside sunspots, and other studies showed that these waves might have been excited beneath sunspots \citep{zha15, cho20}.

Slow magnetoacoustic waves in coronal fan loops are known to damp rapidly \citep[e.g.,][]{dem09}.
Various damping mechanisms have been investigated, such as thermal conduction, compressive viscosity, radiative cooling, and magnetic field divergence \citep[e.g.,][]{ofman00,nak00,dem03,dem04,klim04}.
Among these, thermal conduction has been considered the dominant factor.
Recent analysis using SDO/AIA observations has revealed that the damping of slow magnetoacoustic waves is frequency-dependent \citep{kri14}, and the frequency-dependent damping ($\tau \sim P^a$) can be explained by the damping of slow waves due to thermal conduction ($a=0-2$) and compressive viscosity ($a=2$) \citep[][and references therein]{wan21}.
Recently, \citet{kol22} proposed to explain the observed scaling of the damping time with period of standing slow waves based on the mechanism of wave-induced thermal misbalance.

Earlier theories based on models involving magnetic cylinders filled with uniform plasma predicted that the slow waves are highly dispersive.
Recent theoretical investigations suggested that the dispersion properties of slow waves can be significantly influenced by the misbalance between plasma cooling and heating processes induced by the waves in the corona \citep{zav19, bel21}.
These dispersive effects may hold the key to understanding the evolution of slow waves in coronal fan loops and explaining the observed frequency-dependent damping behavior.

Although the dominant period of the slow \mac waves is near 3.0\,min (or near 5.6\,mHz in frequency), its period covers a wide range from about 0.5\,min to over 30\,min (or from about 0.05 to over 30\,mHz in frequency), and this is why frequency-dependent damping could be studied previously \citep[e.g.,][]{kri14}.
But whether the phase velocities of the slow waves show frequency dependency is not yet well studied.
It is well known that the $p$-mode waves in the solar photosphere are dispersive because of the high atmospheric stratification \citep{chr02}, but whether the slow \mac waves in the chromosphere and corona, driven by the dispersive photospheric waves, are dispersive has not been studied in literature.
In this paper, through measuring the phase velocities in the Fourier domain, we demonstrate that the slow magnetoacoustic waves traveling in coronal fan loops are dispersive, with phase velocities varying substantially with the waves' frequency.
This paper is organized as follows: in Section~\ref{sec2} we introduce the data and analysis method, and present our analysis results. 
In Section~\ref{sec3}, through a couple of measurement experiments we demonstrate our analysis is robust and trustworthy.
We then discuss the results and give conclusions in Section~\ref{sec4}.

\section{Data, Analysis, and Results}
\label{sec2}

\subsection{Analysis and Results from \sdo/AIA Data}
\label{sec21}

\subsubsection{\sdo/AIA Data and Waves}
\label{sec211}

\begin{figure*}
\centering
\includegraphics[width=0.9\textwidth]{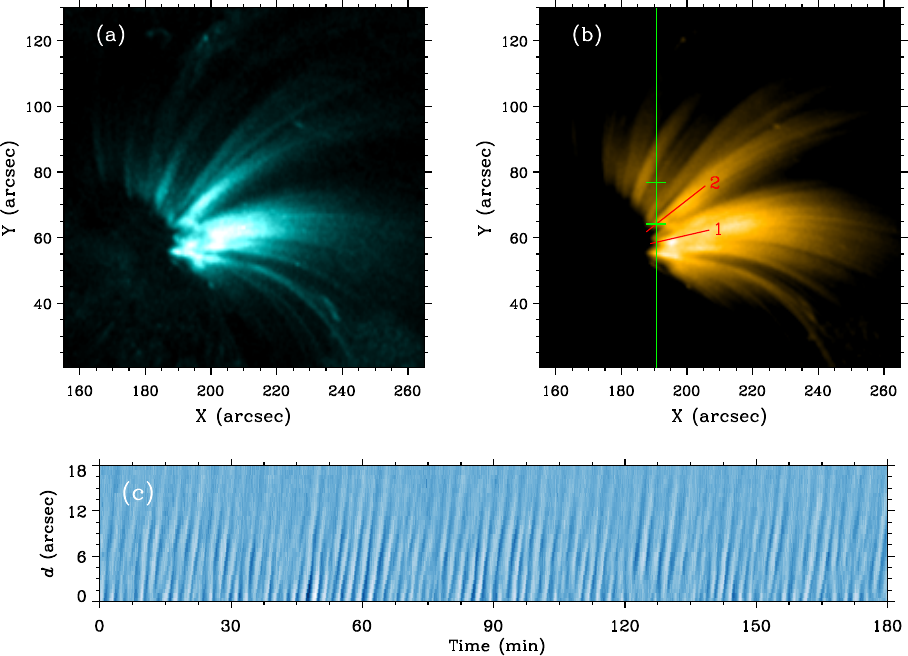}
\caption{(a) \sdo/AIA 131\,\AA\ image of the analyzed area.
(b) \sdo/AIA 171\,\AA\ image of the analyzed area. 
Red lines 1 and 2 indicate along which our computation of phase velocities are made. 
The vertical green line indicates the position of the \hinode/EIS slit, and the two  horizontal bars mark the boundaries between which the analysis presented in Section~\ref{sec23} is carried out.
(c) Time--space diagram along Line 1 in 171\,\AA\ for the entire time sequence showing the propagation of waves. Location $d$ measures from the left-most point of Line 1.
The accompanying animation shows wave propagation in 171\,\AA. } 
\label{img_171}
\end{figure*}

In this study, we select the coronal fan loops above the sunspot in NOAA Active Region~11899 for analysis.
The analysis uses 3 hours of \sdo/AIA observations, covering the period from 12:00 to 15:00\,UT of 2013 November 19 when the region was close to the disk center. 
It should be kept in mind that the velocities measured in this study are a plane-of-sky projection of the propagation velocities of the studied waves.
Data from all the six \sdo/AIA EUV wavelength channels (including 335\,\AA, 211\,\AA, 193\,\AA, 171\,\AA, 131\,\AA, and 94\,\AA), which have a temporal cadence of 12\,s and a pixel size around $0.6\arcsec$,  are obtained and analyzed.
Figure~\ref{img_171} exhibits snapshots of the analyzed area in 131\,\AA\ and 171\,\AA\ channels, which show similar but distinctly different loop structures.
Based on the temperature response functions, these two wavelength channels are primarily sensitive to the temperatures of $10^{5.6}$\,K and $10^{5.8}$\,K, respectively \citep{boe12}.

The propagating intensity oscillations, with characteristic periods around and below 5 min, observed in these \sdo/AIA data (see the animation created using 171\,\AA\ data, associated with Figure~\ref{img_171}) are interpreted as magnetoacoustic waves (see Section~\ref{sec1}), primarily due to their phase speeds being close to the sound speed in coronal loops. The phase speeds, mostly below 100\,\kms (see Section~\ref{sec213}), are inconsistent with typical kink or Alfv\'{e}nic waves, which have speeds of $1000 - 2000$\,\kms\ \citep{kid12, dem12, wan16}.
Previous observational and theoretical studies suggest that weak intensity oscillations may be caused by kink waves due to variations in the LOS depth of loops \citep[e.g.,][]{coo03}. However, for kink waves with periods of $3-5$ min, the wavelengths are of several hundred Mm long and their LOS-depth variations are negligible.
It is also possible that transverse oscillations of coronal loops alternately shift in and out of the slit lines selected in our analysis (see Section~\ref{sec212}), creating the wave features we analyze. However, the amplitude of transverse oscillations in coronal loops is typically $ < 5$ \kms\ \citep{mci11, wan16}, making their effect negligible in our analysis.

\subsubsection{Analysis Method}
\label{sec212}

We compute the phase velocities as functions of frequency along Lines 1 and 2 (Figure~\ref{img_171}b), selected near the middle line of their corresponding coronal loops, where slow \mac waves are clearly visible (see Figure~\ref{img_171}c). 
It should be noted that the \mac waves may or may not propagate exactly along the selected lines. 
However, this analysis concerns more the frequency dependency of the wave phase velocities rather than the propagation velocities themselves, therefore whether the selected lines are in agreement with the propagation path does not impact our analyses or conclusions.
To enhance the waves signals, we perform detrending by removing from each image a running background averaged over 16\,min.

To compute the phase velocities along Line 1, we follow the Fourier analysis method outlined by \citet{che18}. 
Two time sequences, $\psi(d,t)$ at location $d$ on Line 1 and $\psi(d+\Delta,t)$ at a distance of $\Delta$ to the right (or downstream of the waves) of $d$ along Line 1, are taken and the cross-correlation function between the two sequences are computed in the Fourier domain following
\begin{equation}
R(d, \Delta,\nu) = \widehat{\psi} (d, \nu) \cdot \widehat{\psi}^{\dagger} (d+\Delta, \nu),
\label{eq1}
\end{equation}
where \ $\widehat{}$ \ represents applying Fourier transform on time (thus, time $t$ becomes frequency $\nu$), $^\dagger$ represents taking complex conjugate, and $R(d, \Delta,\nu)$ is a complex number corresponding to each $d, \Delta$, and $\nu$.
For each $\Delta$, one set of $R(d, \Delta, \nu)$ is obtained with $d$ running from the left end of Line 1 to the right end and $\nu$ ranging from 0 to Nyquist frequency. 
Technically, one set of phase velocities, as a function of $d$ and $\nu$, can be computed from $R(d, \Delta, \nu)$ along Line 1 but the signal-to-noise ratio is low.
To enhance the signal-to-noise ratio, we repeat the calculations along four lines parallel to Line 1, two lines above and two lines below with distances of 1 and 2 pixels from Line 1 separately, and then average the five sets of $R(d, \Delta,\nu)$'s assuming all these five lines have the same wave properties.
The $R(d, \Delta,\nu)$ has a frequency step of 93\,$\mu$Hz, which is more than necessary in our analysis, therefore we bin down every 5 frequency steps into one. 
We further smooth the $R(d, \Delta,\nu)$ by 3 pixels in both $d$ and $\nu$, and the resultant cross-correlation function is denoted as $\langle R(d, \Delta, \nu) \rangle$.

The $\langle R(d, \Delta, \nu) \rangle$ carries information of phase shifts at the position $d$ as a function of $\nu$ between two locations $\Delta$ apart, and the phase shifts can be calculated following
\begin{equation}
\delta \phi (d, \Delta, \nu) = \arg \Bigl( \langle R(d, \Delta, \nu) \rangle \Bigr),
\label{eq2}
\end{equation}
where $\arg$ represents computing the phase angle of the complex number.
For each $\nu$ this $\delta \phi$ represents a travel time of $\delta t = \delta \phi / (2 \pi \nu)$ between the two locations, and the phase velocities at $d$ is therefore
\begin{equation}
V_\mathrm{ph} (d,\nu) = \frac{2\pi\nu\Delta}{\delta\phi(d, \Delta, \nu)}.
\label{eq3}
\end{equation}

\begin{figure*}
\centering
\includegraphics[width=0.80\textwidth]{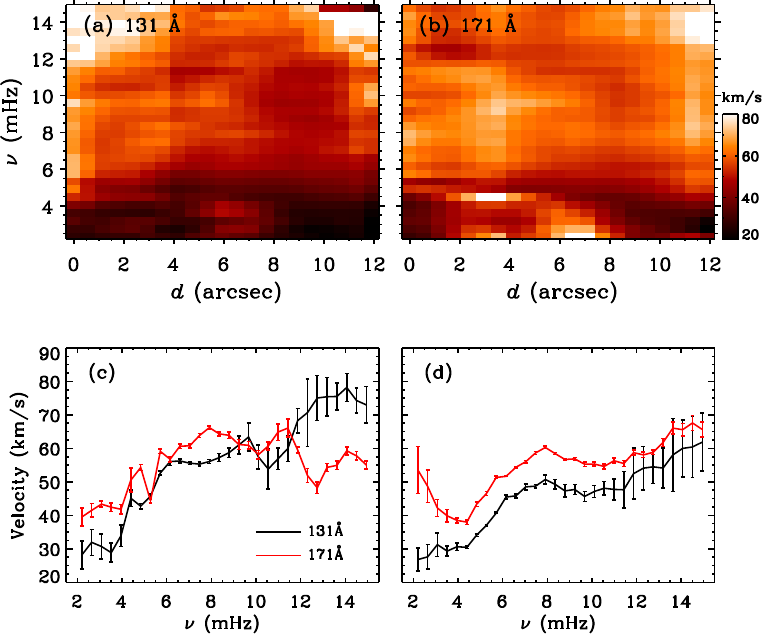} 
\caption{(a) Measured phase velocities along Line 1 using \sdo/AIA 131\,\AA\ data, displayed as a function of location $d$ and frequency $\nu$. 
White pixels in the panel are where computation of phase velocities fails.
(b) Same as panel (a) but for 171\,\AA\ observations.
(c) Comparison of phase velocities obtained from 131\,\AA\ and 171\,\AA\ data at $d = 3.0\arcsec$, displayed as functions of $\nu$.
(d) Same as panel (c) but for $d= 10.2\arcsec$.}
\label{velo_line1}
\end{figure*}

In our calculations, we choose $\Delta$ to be from 1 to 5 pixels, thus a total of five sets of $V_\mathrm{ph} (d,\nu)$ is obtained corresponding to five $\Delta$'s.
Our results show that each set of $V_\mathrm{ph} (d,\nu)$ is in good agreement with the others, demonstrating the robustness and reasonableness of our analysis.
These five sets of measurements are then averaged for the final phase velocities $V_\mathrm{ph} (d,\nu)$, and the errors are also estimated for each location and frequency.
The computation of phase velocity is thought failed if at a certain $(d, \nu)$, the derived speed is negative, or is extremely large with values above 200\,\kms, or the error bar exceeds the value of the speed.

\subsubsection{Analysis Results}
\label{sec213}

Figure~\ref{velo_line1} shows the phase velocities derived along Line 1 using \sdo/AIA 131\,\AA\ and 171\,\AA\ data, in two dimensions as functions of $d$ and $\nu$, as well as in one dimension as functions of $\nu$ for selected $d$'s.
It can be seen from our measurements that the phase velocities clearly vary with frequency, generally increasing with frequency but the trend is not monotonic.
The phase velocities vary between speeds of 30 and 80\,\kms. 
Around the frequency with dominant wave power (about 6\,mHz with 3\,min period), the phase velocity is about $40-50$\,\kms, and this is in agreement with the past measurements without the frequency decomposition \citep[e.g.,][]{kri17}. 

\begin{figure*}
\centering
\includegraphics[width=0.80\textwidth]{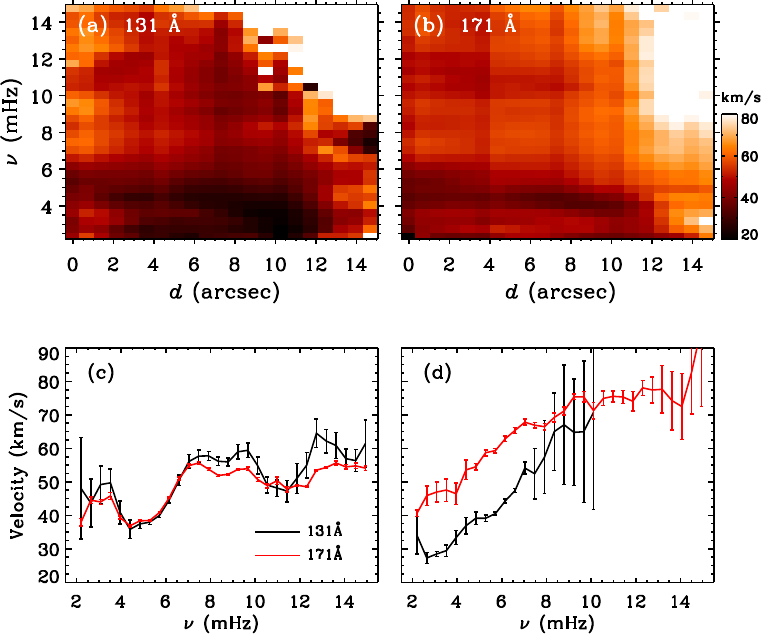} 
\caption{(a) Measured phase velocities along Line 2 using \sdo/AIA 131\,\AA\ data, displayed as a function of $d$ and $\nu$. 
White pixels in the panel are where computation of phase velocities fails.
(b) Same as panel (a) but for 171\,\AA\ observations.
(c) Comparison of phase velocities obtained from 131\,\AA\ and 171\,\AA\ data at $d= 1.2\arcsec$, displayed as functions of $\nu$.
(d) Same as panel (c) but for $d = 9.6\arcsec$.}
\label{velo_line2}
\end{figure*}

Phase velocities along Line 2 is obtained following the same analysis process, and Figure~\ref{velo_line2} shows the results.
These phase velocities also show clear frequency dependency, with a general trend of velocities increasing with frequency.
These trends are qualitatively similar to those along Line 1, demonstrating that this property of waves --- phase velocities are wave frequency dependent --- are ubiquitous rather than due to specific location choices.
It can also be noticed that the phase velocities measured from 131\,\AA\ have similar values with those measured from 171\,\AA\ near the footpoints (Figures~\ref{velo_line1}c and \ref{velo_line2}c), but are substantially smaller in the higher altitude (Figures~\ref{velo_line1}d and \ref{velo_line2}d).
This may due to that the 171\,\AA\ channel is more sensitive to higher temperatures ($10^{5.8}$\,K) than the 131\,\AA\ channel ($10^{5.6}$\,K) in this area.

Are these frequency-dependent phase-velocity measurements robust, or they are just artifacts due to noises? 
To answer this question, in Section~\ref{sec3}, we carry out extra measurements and data experiments to demonstrate that the phase-velocity measurements from our analysis technique are robust and trustworthy, not artifacts from noises or due to analysis procedures.
And, it is also not surprising that the slow waves have frequencies up to 15\,mHz. 
The slow waves likely originate from the photosphere and chromosphere, where waves with frequencies above 5.5\,mHz (cut-off frequency with vertical magnetic fields) can travel through. 
Waves with frequencies lower than 5.5\,mHz may propagate into the corona along inclined magnetic fields \citep{dep05, wan09}. 
Local perturbations in the corona and coronal loops can also generate waves of various frequencies.

\subsection{Analysis and Results from DEM Temperature Maps}
\label{sec22}

To further examine the frequency dependency of the slow \mac waves' phase velocities, we use another set of data --- differential emission measure (DEM) maps.
Since individual coronal loops are typically isothermal due to high thermal conductivity \citep{lan08, lan10}, the overlap of structures with different temperatures may result in a broad DEM distribution. 
Thus, the DEM analysis may help decompose different coronal structures along the line-of-sight direction.

The DEM code developed by \citet{che15} is applied on the same sets of \sdo/AIA observations used for analysis in Section~\ref{sec21}, and maps of temperatures ranging from $10^{5.5}$ to $10^{7.5}$\,K are derived.
However, since the temperatures of coronal fan loops typically range in $0.6 - 1.0$\,MK \citep{del03, you07, bro11}, we believe only the DEM for temperatures below $10^{6.0}$\,K are reliable in this region for our further analysis, while the DEM for temperatures above that may be due to cross talk with low temperatures.

\begin{figure*}
\centering
\includegraphics[width=0.90\textwidth]{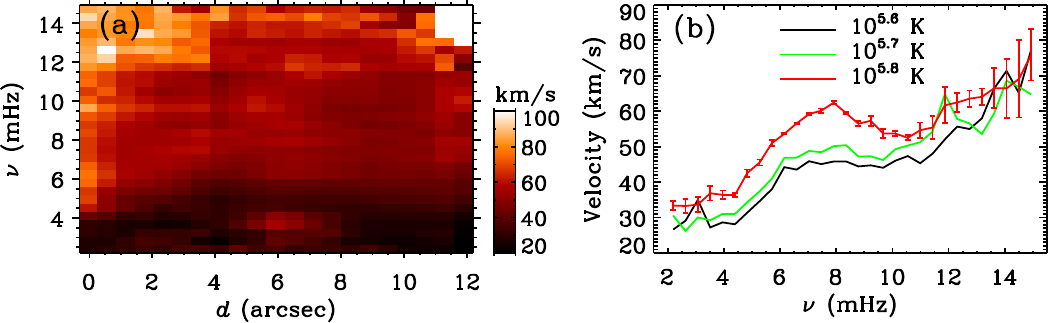} 
\caption{(a) Phase velocities, obtained along Line 1 from the DEM maps at temperature of $10^{5.7}$\,K and displayed as a function of location $d$ and frequency $\nu$.
White pixels in the panel are where computation of phase velocities fails.
(b) Phase velocities, taken from three different DEM temperatures at $d = 9.6\arcsec$, are displayed as functions of $\nu$. 
For clarity, error bars are only plotted in one of the velocity curves but are similar in all three.}
\label{dem_res}
\end{figure*}

We then apply the same measurement procedure used in Section~\ref{sec21} on these DEM temperature maps, and derive the phase velocities along Lines 1 and 2 for each DEM temperature as functions of location and frequency.
Figure~\ref{dem_res}a displays the phase velocities calculated along Line 1 for $T_\mathrm{DEM} = 10^{5.7}$\,K.
It can be seen, similarly to the results in Figures~\ref{velo_line1} and \ref{velo_line2}, the phase velocities are highly frequency dependent, varying substantially with the frequency.

Figure~\ref{dem_res}b shows the phase velocities, as functions of frequency, for different DEM temperatures at one selected location $d= 9.6\arcsec$.
It can be seen that for all the temperatures, the phase velocities show a similar trend of increase with frequency but with apparent fluctuations.
Generally speaking, the phase velocities increase with the $T_\mathrm{DEM}$, but this is not strictly true everywhere in this set of data.
Whether the phase velocities follow a trend of $v_\mathrm{ph} \propto \sqrt{T}$ \citep{uri13} is not immediately clear in this analysis.

\subsection{Analysis and Results from \hinode/EIS Data}
\label{sec23}

Our third set of analysis uses observations from the EUV Imaging Spectrometer \citep[EIS;][]{cul07} onboard {\it Hinode}.
The data used in this analysis were acquired through EIS Study 434 from the same active region analyzed in Sections~\ref{sec21} and \ref{sec22} during approximately the same period from 12:00 to 15:08\,UT, 2013 November 19.
The observations, using a few spectral lines, were obtained along a $2\arcsec$-wide and $512\arcsec$-long slit working in a sit-and-stare mode (i.e., the slit is placed at a fixed solar location) with a cadence of approximately 31.4\,s. 
In this study, we analyze the Doppler velocities inferred from the spectral line, Si VII 275.35\,\AA, which is formed at, approximately, a plasma temperature of $10^{5.8}$\,K, similar to the \sdo/AIA 171\,\AA\ channel.
We correct the pointing position of the \hinode/EIS slit by coaligning its raster images in \ion{He}{2} 256.32\,\AA\ observed during 10:40 -- 11:59\,UT with an \sdo/AIA 304\,\AA\ (\ion{He}{2}) image at 10:40\,UT, with an accuracy of about $1-2\arcsec$.
The position of the \hinode/EIS slit is marked in Figure~\ref{img_171}b, and the slit does not align with any coronal loops hence unlikely corresponds to any wave propagation direction.

\begin{figure*}
\centering
\includegraphics[width=0.90\textwidth]{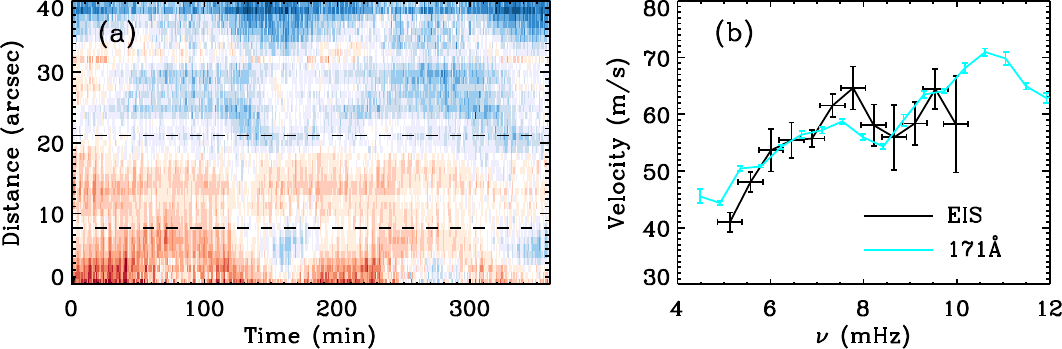}
\caption{(a) Doppler velocities obtained from line \ion{Si}{7} 275.35\,\AA\ along the \hinode/EIS slit, with the red (blue) indicating redshift (blueshift) and the display ranging from $-20$ to $+20$\,\kms.
The displayed vertical range is only part of the slit (see Figure~\ref{img_171}b).
The two horizontal dashed lines delimit the locations where phase velocities are measured.
(b) Comparison of phase velocities measured from the \hinode/EIS Doppler data and \sdo/AIA 171\,\AA\ intensity data, displayed as functions of frequency $\nu$. 
Horizontal error bars for the dark curve indicate the frequency range within which the data points are averaged. }
\label{eis}
\end{figure*}

Figure~\ref{eis}a shows the line-of-sight Doppler velocities obtained along the \hinode/EIS slit as a function of time.
These data along the slit can be analyzed essentially in the same manner as those \sdo/AIA data along Lines 1 and 2, except that they do not have observations in parallel lines to average for a better signal-to-noise ratio.
Before these Doppler data are used for the cross-correlation calculation in Fourier domain, at each time step we remove a running average over 16 minutes, and at each position along the slit we remove a running average of 5 pixels.
The former removal is the same as in our \sdo/AIA data analysis in Section~\ref{sec21}, while the latter is not done in the AIA data but is necessary for the EIS data (see below). 

We follow the same steps prescribed in Section~\ref{sec21}, and compute the phase velocities $V_\mathrm{ph} (d, \nu)$ from the position $8\arcsec$ to $21\arcsec$, beyond which range the slow \mac waves are not clearly visible. 
The distance $\Delta$ used in this calculation includes 1, 2, 3, and 4 pixels, whereas one pixel has a size of $1\arcsec$. 
If not removing the 5-pixel-wide average along the slit, the calculated phase velocities for the case of $\Delta=1$ pixel, i.e., between neighboring pixels, are a few times faster than those calculated from the cases of $\Delta=2 - 4$; however, the results become more consistent for different $\Delta$'s after the removal of the running average. 
This indicates that the neighboring pixels have a tendency of simultaneous variations, implying a large point spread function in the \hinode/EIS observations.
This point spread function may be as large as $5\arcsec$, consistent with what was found by \citet{you22}.

After averaging the phase speeds measured from all $\Delta$'s, we further average the results of all positions $d$ to enhance the signal-to-noise ratio. 
The frequency step for this set of data is 88\,$\mu$Hz, and we again average every 6 frequency steps.
For comparison, \sdo/AIA 171\,\AA\ data from the identical EIS slit position is taken, and the exactly same measurement procedure are applied to derive the phase velocities. 
Results from these two sets of data are shown in Figure~\ref{eis}b.
It can be seen that the phase velocities measured from both the \hinode/EIS Doppler velocities and \sdo/AIA 171\,\AA\ intensities are in reasonable agreement.
Both results show that the phase velocities of the slow \mac waves are highly frequency dependent, and the speeds generally increase with wave frequency.

\section{Robustness of the Measurements }
\label{sec3}

To examine how robust our measured dispersiveness of the slow \mac waves is, particularly in the high-frequency regime, we carry out two sets of measurement experiments (Sections~\ref{sec31} \& \ref{sec32}) and three sets of artificial data experiments (Section~\ref{sec33}).

\begin{figure*}
\centering
\includegraphics[width=0.80\textwidth]{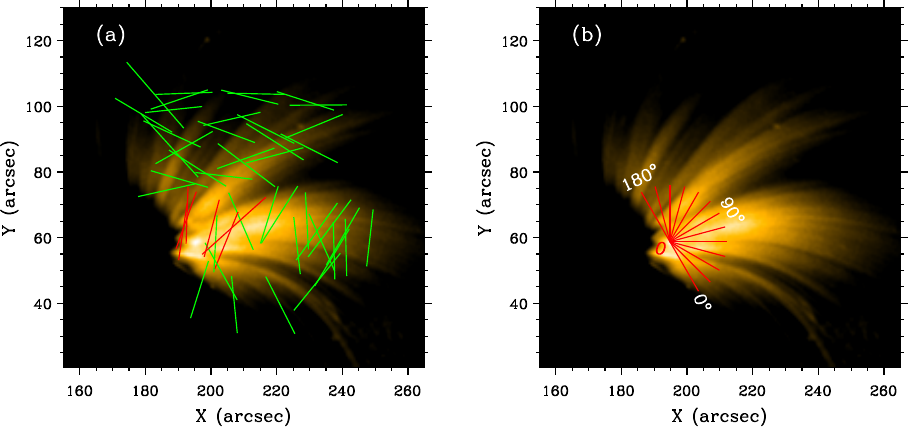}
\caption{(a) Green and red lines overplotting the 171\,\AA\ image indicate the 50 randomly selected lines along which our analysis procedure is applied. 
Our phase-velocity measurements along the green lines shows salt-and-pepper results, and the summary of these measurements is given in Section~\ref{sec31}.
However, part of the measurements along the red lines show reasonable values in phase velocities, and we therefore design a new set of measurements that are reported in Section~\ref{sec32}.
(b) Thirteen red line segments, also overplotting the 171\,\AA\ image, show the lines along which phase velocities are measured to examine the azimuthal variations of the measured phase velocities (see Section~\ref{sec32}).  }
\label{image_lines}
\end{figure*}

\subsection{Measurements along Random Lines}
\label{sec31}

One may argue that the dominant power in the slow \mac waves is around 6.0\,mHz, and at frequencies far from this dominant frequency are just noises, and therefore, the phase velocities measured at those frequencies are also noises thus untrustworthy. 
To examine this argument, we select 50 lines randomly in the field of view, all of which pass the bright coronal area crossing but not along the coronal loops (see Figure~\ref{image_lines}a). 
We then follow the procedure described in Section~\ref{sec21} and calculate cross-correlation functions at all positions along these lines, from which power and phase speed are calculated at different frequencies for each line. 

\begin{figure*}
\centering
\includegraphics[width=0.80\textwidth]{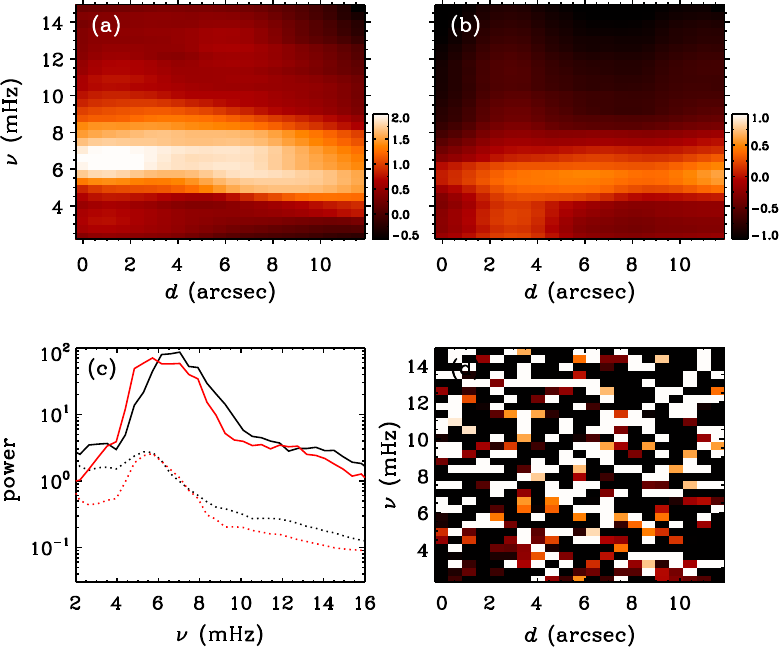}
\caption{(a) Power of the cross-correlation functions, shown as functions of line position $d$ and frequency $\nu$, measured along Line~1 in 171\,\AA, and displayed in the logarithm scale. 
(b) Power of the cross-correlation functions, averaged from the measurements along the 44 randomly selected green lines shown in Figure~\ref{image_lines}a, and displayed in the logarithm scale as well.
Note that panels (a) and (b) are displayed in different color scales. 
(c) Comparing the power of the cross-correlation functions, as functions of frequency, obtained from Line~1 (solid curves) and averaged from the random lines (dotted curves), at the positions of $d = 3.6\arcsec$ (dark curves) and $8.5\arcsec$ (red curves). 
(d) Phase velocities, displayed as functions of $d$ and $\nu$, obtained from one of the 44 randomly selected lines. 
Results from the rest of the lines are similar.
}
\label{noise_power}
\end{figure*}

Examining the 50 sets of two-dimensional phase velocities as functions of $\nu$ and $d$, shown in a way like Figure~\ref{velo_line1}a, we find that all but 6 sets of them show random salt-and-pepper-like values, as shown in Figure~\ref{noise_power}d. 
Therefore, in this Section~\ref{sec31}, we only further analyze the results from the 44 lines (marked green in Figure~\ref{image_lines}a), and will discuss those 6 lines (marked red in Figure~\ref{image_lines}a) in Section~\ref{sec32}. 

Figure~\ref{noise_power}a, b, and c show the comparison of the power from Line 1 (see Section~\ref{sec2}) with that averaged from the 44 random lines. 
It can be found that the averaged power along the random lines shows a frequency-dependent power distribution similar to that from along the coronal loop, but over 1 order of magnitude weaker. 
This is not surprising, because all these random lines pass coronal loops, picking up the oscillations in the loops, which have a dominant power around 6.0\,mHz and relatively weaker power outside the range of $5 - 8$\,mHz.
The phase velocities calculated from these 44 random lines all show random salt-and-pepper-like values (Figure~\ref{noise_power}d), indicating that oscillatory signals along these random lines do not possess coherent phase relations. 

This set of measurements demonstrate that no reasonable phase velocities can be measured out of observational noises, in either dominant frequencies or in weak-power frequencies.
The fact that consistent phase velocities are measured at weaker-power frequencies along coronal loops in Section~\ref{sec2} supports that those measurements are not noises.

\begin{figure*}
\centering
\includegraphics[width=0.80\textwidth]{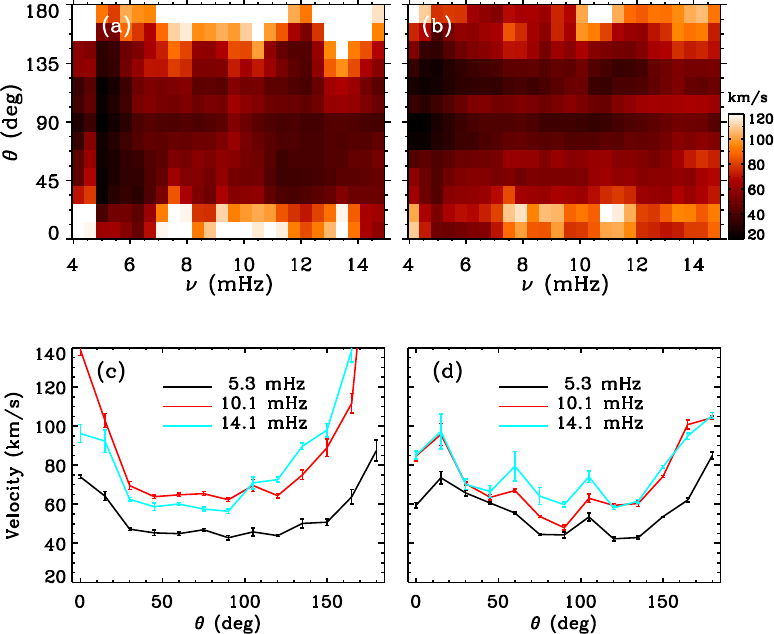}
\caption{(a) Phase velocities measured at the position $d = 1.2\arcsec$, displayed as functions of frequency $\nu$ and azimuthal angle $\theta$. 
White areas in the panel indicate where our calculations do not converge. 
(b) Same as panel (a) but at the position $d = 5.5\arcsec$. 
(c) Comparison of the phase velocities, taken from panel (a) at different $\nu$'s and shown as functions of $\theta$.
Dark, red, and cyan curves are for $\nu = 5.3, 10.1$ and $14.1$\,mHz, respectively.
(d) Same as panel (c) but curves are taken from panel (b).  } 
\label{noise_velo}
\end{figure*}

\subsection{Measurements with Different Angles Relative to Propagation Direction}
\label{sec32}
Examining the two-dimensional phase velocities measured along the 6 red lines in Figure~\ref{image_lines}a, we find that a fraction (large or small) of the velocity images show reasonable values, with values falling between 0 to 200\,\kms\ consistently across multiple pixels in both $\nu$ and $d$. 
These lines locate close to the footpoints of the coronal loops, and the phase velocities measured from these lines could be components of the phase velocities of the slow \mac waves.
To further investigate this, we carry out a second set of measurements for a systematic examination of how the measured phase speeds vary with the azimuthal angle around the waves' propagation direction.

As shown in Figure~\ref{image_lines}b, we choose one spot, `$O$', on Line~1 (see Section~\ref{sec2}) together with 13 lines, which have one end on `$O$' with azimuthal angles uniformly spanning from $0\degr$ to $180\degr$.
The same measurement procedure is then applied on these 13 separate lines, and each line returns with phase velocities as functions of frequency and positions. 
All or most of these measured phase velocities fall in the reasonable speed range of $ 0 - 200$\,\kms.
Figure~\ref{noise_velo}a and \ref{noise_velo}b show the measured phase velocities at the positions $d = 1.2\arcsec$ and $5.5\arcsec$ away from spot `$O$', as functions of azimuthal angle $\theta$ and frequency $\nu$. 
Figure~\ref{noise_velo}c and \ref{noise_velo}d display comparisons of the phase velocities as functions of $\theta$ for selected $d$'s and $\nu$'s.

\begin{figure}
\centering
\includegraphics[width=0.35\textwidth]{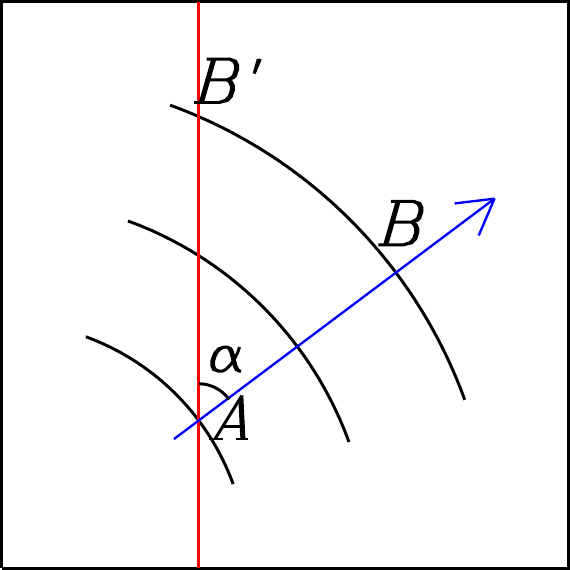}
\caption{Schematic plot showing phase velocities measured along different directions can result in different values.
Line $AB$ is along the wave's propagation direction, while line $AB^\prime$ has an angle $\alpha$ relative to $AB$.
Cross-correlation analysis is expected to give similar travel times from $A$ to $B$ and from $A$ to $B^\prime$ (because $B$ and $B^\prime$ have a same phase), but $AB^\prime$ is longer than $AB$ in distance, resulting in a faster phase-velocity measurement. }
\label{schematic}
\end{figure}

As the schematic plot in Figure~\ref{schematic} shows, phase velocities measured at different azimuthal angles relative to the waves' propagation direction can be quite different, being smallest when $\alpha = 0\degr$ and largest when $\alpha$ is close to $\pm90\degr$, where $\alpha$ is the angle between the wave propagation direction and the measurement direction. 
Our measurements shown in Figure~\ref{noise_velo} nicely exhibit this expected variation of the measured velocities with the azimuthal angle. 
More notably, this azimuthal angle dependency is not only seen in the phase velocities measured at the dominant frequencies, but is also seen in the weak-power frequencies far away from the dominant frequencies, such as $\nu=14.1$\,mHz (Figure~\ref{noise_velo}c and \ref{noise_velo}d). 
If the phase velocities measured at the weak-power frequencies are noises or are unrelated to the slow waves, then they are not expected to show such a dependency on the azimuthal angle. 
This observational fact presents another piece of evidence that the phase velocities measured at frequencies with weaker power, as presented in Section~\ref{sec2}, are trustworthy. 

It is worth noting that Figure~\ref{schematic} presents only a simplified geometry in which waves propagate, and several other factors may contribute to the complexity of the measured results. 
Figure~\ref{schematic} assumes the waves have a circular wavefront; however, since temperature, density, and magnetic field vary with the angle relative to the coronal loops, the wave speed can also vary with the propagation angle.
Moreover, without knowing the wave-source location, the lengths of $AB$ and $AB^\prime$ may or may not differ significantly, meaning the phase speeds measured along different directions may not vary substantially. 
This explains why the phase speeds measured in Figure~\ref{eis}b fall within a range similar to those measured along Lines 1 and 2. 
It is also worth pointing out that the speeds measured in Figure~\ref{eis}b correspond approximately to those at an azimuthal angle of $150\degr$ in Figure~\ref{noise_velo}c.


\subsection{Testing Power and Phase Leakage in Fourier Analysis}
\label{sec33}

Another concern regarding the analyses of frequency-dependent phase speeds, as presented in Section~\ref{sec2}, is that the power at frequencies outside $4.8 - 9.0$\,mHz is weak (see Figure~\ref{noise_power}c), and one may suspect that this is due to the leakage of the dominant power during the Fourier analysis applied on a data sequence with limited time duration and certain temporal cadence. 
This broadening effect may cause miscalculations of phase shifts, thus the phase speeds, at weak-power frequencies. 

First, we stress that the power being weak at some frequencies does not necessarily mean the power and phases calculated at these frequencies are not trustworthy. 
For instance, in the solar photosphere, the oscillatory power around $6.0 - 7.0$\,mHz is 2 to 3 orders of magnitude weaker than the peak power near 3.0\,mHz, but those power is not only reliable but also is very useful in studying the solar interior \citep{chr02}. 
On the other hand, we also agree that convincing evidence is needed to prove that the power and phase shifts calculated in this study outside the $5 - 8$\,mHz range are not due to leakage from the dominant power in the process of Fourier analysis.
Below, we present two separate data experiments using Monte Carlo method to investigate whether the Fourier analysis contaminates our analysis.

\begin{figure*}
\centering
\includegraphics[width=1.0\textwidth]{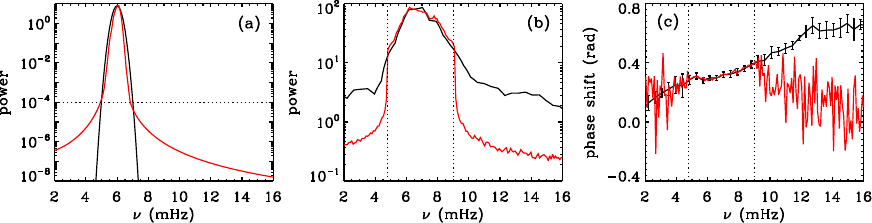}
\caption{(a) Comparison of the prescribed power distribution (dark) and the distribution calculated from Monte-Carlo simulated data sequences (red) for the narrow-band case. 
The dotted line at $10^{-4}$ (or $10^{-5}$ of the peak power) indicates where the Fourier-analysis-caused broadening starts. 
(b) Comparison of the observed power distribution (dark) and the the distribution calculated from Monte-Carlo simulated data sequences (red) for the wide-band case. 
The simulation only takes the observed power between the two vertical dotted lines. 
(c) Comparison of phase shifts from observational measurements (dark curve with error bars) and from the Monte Carlo simulated data with wide frequency band (red curve). 
The two vertical dotted lines indicate the frequency range where measured phase shifts are embedded in the simulation. }
\label{monte_carlo}
\end{figure*}

\subsubsection{Narrow Band Test without Noises}
\label{sec331}

The first data experiment examines whether the Fourier analysis broadens a narrow-band power distribution without considering noises. 
We first generate an artificial power distribution $P(\nu)$, which is a Gaussian function with a FWHM of 0.5\,mHz peaking at 6.0\,mHz. 
Then we construct a time sequence that has the same power distribution $P(\nu)$, the same time duration, and the same temporal cadence as the data used in this study, i.e., 3 hr with 12 sec cadence. 
The sequence is constructed following the steps below:

\begin{enumerate}
\item The constructed sequence is a sum of numerous monochromatic sequences, each of which has the same duration and cadence with a frequency falling between 4.0 and 8.0\,mHz, i.e., 2.0\,mHz wide on either side of the peak.
\item At each frequency, the amplitude of the monochromatic wave is equal to $\sqrt{P(\nu)}$. 
\item For each frequency, the phase of the monochromatic wave is randomly generated. 
\end{enumerate}

Because the power of cross-correlation functions is used in this study, we need to construct another time sequence that can be used to cross-correlate with the sequence we just construct. 
To construct this second sequence, we adopt the same process as described above but with a phase different from the first sequence at each frequency, and the phase difference is also a random value between 0.0 and 1.0\,rad. 
Then, a cross-correlation is computed between these two sequences in Fourier domain following Equation~\ref{eq1} without the $d$ and $\Delta$ parameters, and the power of the cross-correlation function is computed as a function of $\nu$. 

The above procedure is then repeated 100 times with random realizations, and the mean power from these 100 realizations is calculated. 
The comparison of the prescribed power distribution and the calculated distribution in Fourier domain (Figure~\ref{monte_carlo}a) shows that Fourier analysis does not broaden the power distribution until the power is less than $10^{-5}$ of the peak power. 
Such a broadening is totally negligible in the observational analysis, as $10^{-5}$ is much smaller than observational uncertainties. 

\subsubsection{Wider Band Test with Noises and Phase Shifts}
\label{sec332}

The first experiment demonstrates that Fourier analysis does not broaden the power range of a noiseless  and narrow-frequency-band sequence.
In the second data experiment, we examine whether Fourier analysis broadens a wider power distribution, with noises, and with relative phase shifts. 
The first step of our data-preparation procedure is similar to the previous experiment except that we use the observed power distribution (dark curve in Figure~\ref{noise_power}c) rather than an artificial one. 
When constructing the time sequences, we choose the frequency range between 4.8 and 9.0\,mHz where the power is above $10^{-2}$ of the peak power.
Relative phase shifts are introduced between the two simulated sequences. 
The frequency-dependent phase shifts are taken from the measurements in Line 2 (see Section~\ref{sec2}) and are shown as the dark solid line with error bars in Figure~\ref{monte_carlo}c. 
Then, two sets of three-hour-long white noises are generated and added to each of the sequences. 
The white noises follow a normal distribution with their average amplitude (defined as the root-mean-square of the total sequence) equal to 20\% of the average amplitude of the signal sequences. 
Note that the noise levels for different \sdo/AIA wavelength channels are also different, but 20\% would be a good approximation in our analysis; however, we also want to point out that a choice of amplitude up to 50\% does not change our conclusion because white noises are spread to all wave frequencies and can be canceled out during the averaging process.
Then, cross-correlation computation is carried out between the two sequences for the frequency-dependent power distribution and phase shifts.

Similar to the first experiment, 100 realizations are carried out, and the mean power distribution and phase shifts are obtained in the Fourier domain (Figure~\ref{monte_carlo}b and \ref{monte_carlo}c). 
Compared to the observed power, 
the simulated power is remarkably similar to the observed power within the range of $4.8 - 9.0$\,mHz, which is expected. 
Outside that range, the simulated power drops rapidly to about one order of magnitude weaker than the observed power. 
This demonstrates that the observed power outside the range of  $4.8 - 9.0$\,mHz is primarily real signal rather than leakage from the peak power, although it may be slightly contaminated by the leakage.

Figure~\ref{monte_carlo}c shows the comparison of phase shifts. 
As expected, the simulated wave sequences can nicely recover the phase shifts embedded between $4.8 - 9.0$\,mHz. 
However, outside that frequency range, the phase shifts become much noisier than the observational measurements. 
This is understandable because the wave in the range is dominated by uncorrelated random noises and the power leakage from the dominant power.
And moreover, the trend of the phase shifts from the simulated data diverges substantially from the observational measurements, particularly for $\nu > 9.0$\,mHz. 
This is another piece of evidence that our measured frequency-dependent phase velocities, inferred from the phase shifts, are not due to signal leakage or noise contamination.

\section{Discussion and Conclusion}
\label{sec4}

Through computing cross-correlation functions in the Fourier domain between time sequences at two locations with certain distances apart, we have calculated the phase velocities of slow \mac waves using three different sets of data: \sdo/AIA intensity data of different wavelength channels, DEM data of different temperatures, and \hinode/EIS Doppler velocity data along a slit. 
All our measurements show that the phase velocities of slow \mac waves exhibit strong dependency on the wave frequency.
The phase velocities generally increase with frequency, but non-monotonic with fluctuations, and the fluctuations are statistically significant.
Since the frequency dependency of the waves' phase speeds is a property of dispersive waves, we therefore conclude that the slow \mac waves traveling in the solar coronal loops are dispersive.

It is understood that the measurements of phase speeds can be contaminated by noises or systematic effects in our analysis procedure, therefore, we carry out extra measurements and data experiments to assess the robustness of our results.
We carry out two extra sets of measurements, one set showing that along random lines in the bright coronal areas, no consistent phase velocities can be measured in both dominant frequencies and in weak-power frequencies despite the fact that their cross-correlation power resembles the lines along coronal loops.
Another set of measurements in the area where slow waves travel past show that along different azimuthal angles, the measured phase velocities show the expected variations in both dominant frequencies and in weak-power frequencies.
Our Monte Carlo data experiments demonstrate that, for the time duration and temporal cadence used in this study, both narrow-band power distribution and wide-band power distribution do not have nonnegligible power leakage into frequencies outside of the dominant power. 
The phase shifts caused by the negligible power leakage are not consistent with the measurements from observations, either.
All these extra measurements and data experiments present strong evidence supporting that the phase velocities measured outside the dominant frequencies are solid and trustworthy with clear and meaningful physics.

It is well known that the temperature response function for each \sdo/AIA wavelength is wide, sensitive to a large range of coronal heights or temperatures of different coronal structures. 
Thus, it may be suspected that different wave frequencies may come from different coronal heights or coronal loops of different temperatures, hence they have different phase velocities. 
However, our analysis using the DEM temperature maps rules out this possibility. 
The DEM maps of various temperatures have largely decomposed the coronal structures of different heights or temperatures, and the phase speeds measured from such maps are expected to fall within a smaller range of heights or temperatures. 
The fact that the results from these DEM temperature maps are essentially similar to those measured from \sdo/AIA intensity images demonstrates that the phase velocities' frequency dependency is a property of the slow \mac waves, but not due to different wave frequencies forming at different coronal heights or temperatures.
And, the good agreement in frequency-dependent phase velocities measured from the \sdo/AIA intensity and the \hinode/EIS Doppler velocity data reinforces this statement, because the Si\,VII line typically forms in a narrow temperature range \citep{you23}.

Our analysis focuses on the plane-of-sky projection of the coronal loops, mostly ignoring the fact that these loops are inclined relative to the perpendicular direction of the solar surface. 
The observed oscillation signals along the coronal loops are likely a LOS combination of signals from loops with slightly different inclination angles. 
Since the cutoff frequency varies with the inclination angle \citep[e.g.,][]{dep05}, it is possible that the observed variations in phase velocities result from contributions of waves traveling in loops with different inclination angles. 
However, as the cutoff frequency generally decreases with the increasing inclination angle \citep{yua14}, the plane-of-sky wave speeds are expected to increase with inclination angle.
Therefore, one would expect the phase speeds to increase with the decreasing wave frequency, opposite to our measurements, and hence the observed dispersiveness does not result from a combination of inclination angle and cutoff frequency. 
Meanwhile, we acknowledge that the extent to which inclination angles vary in the clustered coronal loops and how significantly this effect influences our measured dispersiveness warrant further studies.

Early theoretical studies carried out in the 1980s \citep{edw82, edw83} suggested that slow magnetoacoustic waves traveling along magnetic flux tubes were weakly dispersive due to the finite sizes of their cross sections.
More recently, \citet{bel21} suggested that due to the wave-induced thermal imbalance as well as geometric projection in the line-of-sight direction, the slow magnetoacoustic waves are expected to be dispersive with phase velocities varying with frequency.
However, despite advanced space observations of the UV/EUV coronal loops and a large number of studies on slow magnetoacoustic waves, the dispersiveness of such waves have not been convincingly established observationally.
Our analyses using different observables have demonstrated evidently, for the first time, that the phase velocities of such waves indeed show a strong dependence on the wave frequency, with a trend similar to what was predicted by \citet{bel21} (Cf.~Figures~\ref{velo_line1} -- \ref{eis} of this paper with Figures~3 \& 4 in their paper).
This may indirectly provide observational evidence supporting the presence of heating/cooling imbalance processes in coronal fan loops.
In addition, it is known that a nonlinear non-harmonic wave (e.g. shock-like waves with a steepened front) can lead to dispersion of slow waves in propagation. 
However, since various EUV observations show that these coronal waves are typically of small amplitudes \citep[only a few percentages of background emission, e.g.,][]{dem09}, the nonlinear effect regarding the wave shape is believed to be small. 
But our new results showed in this study suggest that the nonlinearity effect could be an alternative mechanism that needs to be investigated further both observationally and theoretically.

The frequency dependency of the phase velocities in slow magnetoacoustic waves, as pictured in Figures~\ref{velo_line1} -- \ref{eis}, add another layer of information that can be used to diagnose the physical conditions in the coronal loops, in addition to those used in the past studies, such as characteristic frequencies, damping time, and phase velocities.
In their model, \citet{bel21} laid out the theoretical foundation of how we can apply the measurements, as presented in this study, to better determine the physical conditions of the coronal loops, such as magnetic field strengths.

The dispersiveness of slow magnetoacoustic waves may also account for the observed damping of the waves.
The waves of different frequencies propagate with different phase velocities, and after some distance, the wave packet becomes weaker in amplitude and wider in space, causing an apparent damping even if no energy is dissipated.
However, whether this amplitude reduction matches the observed damping and whether this effect can explain the frequency-dependent damping need to be further investigated. 
The dispersiveness of the propagating perturbation of brightness reported in this work may help to distinguish whether these perturbations are magnetoacoustic waves or quasi-periodic upflows.
Considering that speed of flows is unlikely frequency dependent, faster with higher frequency, we believe our analysis results favor the wave nature of these propagating perturbations, just as we assumed at the beginning of our analysis.

In summary, our analyses using three different types of data demonstrate that the slow magnetoacoustic waves traveling in the coronal fan loops are dispersive, exhibiting a general trend of increasing phase velocities with frequency. 
This may help create new opportunities to better determine the physical conditions and parameters in the coronal loops.

\section*{Acknowledgements}
We thank the anonymous referee for constructive comments to help improve the quality of this paper.
\sdo\ is a mission of NASA's Living With A Star program, and AIA is an instrument onboard \sdo, developed and constructed by LMSAL.
\hinode\ is a Japanese mission constructed and launched by JAXA/ISAS, collaborating with NAOJ as a domestic partner, NASA (USA) and PPARC (UK) as international partners.
This work was partly supported by a NASA Heliophysics Guest Investigator grant 80NSSC18K0668.
The work by TW was also supported by NASA grants 80NSSC18K1131, 80NSSC21K1687, and 80NSSC22K0755, as well as a NASA Cooperative Agreement 80NSSC21M0180 to Catholic University of America.

\section*{Data Availability}

The \sdo/AIA data used in our analysis are publicly available at \url{http://jsoc.stanford.edu/}, and the \hinode/EIS data are publicly available at \url{https://solarb.mssl.ucl.ac.uk/SolarB/}.


\bibliographystyle{mnras}
\bibliography{ms} 

\begin{thebibliography}{}
\makeatletter
\relax
\def\mn@urlcharsother{\let\do\@makeother \do\$\do\&\do\#\do\^\do\_\do\%\do\~}
\def\mn@doi{\begingroup\mn@urlcharsother \@ifnextchar [ {\mn@doi@}
  {\mn@doi@[]}}
\def\mn@doi@[#1]#2{\def\@tempa{#1}\ifx\@tempa\@empty \href
  {http://dx.doi.org/#2} {doi:#2}\else \href {http://dx.doi.org/#2} {#1}\fi
  \endgroup}
\def\mn@eprint#1#2{\mn@eprint@#1:#2::\@nil}
\def\mn@eprint@arXiv#1{\href {http://arxiv.org/abs/#1} {{\tt arXiv:#1}}}
\def\mn@eprint@dblp#1{\href {http://dblp.uni-trier.de/rec/bibtex/#1.xml}
  {dblp:#1}}
\def\mn@eprint@#1:#2:#3:#4\@nil{\def\@tempa {#1}\def\@tempb {#2}\def\@tempc
  {#3}\ifx \@tempc \@empty \let \@tempc \@tempb \let \@tempb \@tempa \fi \ifx
  \@tempb \@empty \def\@tempb {arXiv}\fi \@ifundefined
  {mn@eprint@\@tempb}{\@tempb:\@tempc}{\expandafter \expandafter \csname
  mn@eprint@\@tempb\endcsname \expandafter{\@tempc}}}

\bibitem[\protect\citeauthoryear{{Banerjee} et~al.,}{{Banerjee}
  et~al.}{2021}]{ban21}
{Banerjee} D.,  et~al., 2021, \mn@doi [\ssr] {10.1007/s11214-021-00849-0},
  \href {https://ui.adsabs.harvard.edu/abs/2021SSRv..217...76B} {217, 76}

\bibitem[\protect\citeauthoryear{{Belov}, {Molevich}  \&
  {Zavershinskii}}{{Belov} et~al.}{2021}]{bel21}
{Belov} S.~A.,  {Molevich} N.~E.,   {Zavershinskii} D.~I.,  2021, \mn@doi
  [\solphys] {10.1007/s11207-021-01868-4}, \href
  {https://ui.adsabs.harvard.edu/abs/2021SoPh..296..122B} {296, 122}

\bibitem[\protect\citeauthoryear{{Boerner} et~al.,}{{Boerner}
  et~al.}{2012}]{boe12}
{Boerner} P.,  et~al., 2012, \mn@doi [\solphys] {10.1007/s11207-011-9804-8},
  \href {https://ui.adsabs.harvard.edu/abs/2012SoPh..275...41B} {275, 41}

\bibitem[\protect\citeauthoryear{{Brooks}, {Warren}  \& {Young}}{{Brooks}
  et~al.}{2011}]{bro11}
{Brooks} D.~H.,  {Warren} H.~P.,   {Young} P.~R.,  2011, \mn@doi [\apj]
  {10.1088/0004-637X/730/2/85}, \href
  {https://ui.adsabs.harvard.edu/abs/2011ApJ...730...85B} {730, 85}

\bibitem[\protect\citeauthoryear{{Chae}, {Lee}, {Cho}, {Song}, {Cho}  \&
  {Yurchyshyn}}{{Chae} et~al.}{2017}]{cha17}
{Chae} J.,  {Lee} J.,  {Cho} K.,  {Song} D.,  {Cho} K.,   {Yurchyshyn} V.,
  2017, \mn@doi [\apj] {10.3847/1538-4357/836/1/18}, \href
  {https://ui.adsabs.harvard.edu/abs/2017ApJ...836...18C} {836, 18}

\bibitem[\protect\citeauthoryear{{Chen} \& {Zhao}}{{Chen} \&
  {Zhao}}{2018}]{che18}
{Chen} R.,  {Zhao} J.,  2018, \mn@doi [\apj] {10.3847/1538-4357/aaa3e3}, \href
  {https://ui.adsabs.harvard.edu/abs/2018ApJ...853..161C} {853, 161}

\bibitem[\protect\citeauthoryear{{Cheung}, {Boerner}, {Schrijver}, {Testa},
  {Chen}, {Peter}  \& {Malanushenko}}{{Cheung} et~al.}{2015}]{che15}
{Cheung} M. C.~M.,  {Boerner} P.,  {Schrijver} C.~J.,  {Testa} P.,  {Chen} F.,
  {Peter} H.,   {Malanushenko} A.,  2015, \mn@doi [\apj]
  {10.1088/0004-637X/807/2/143}, \href
  {https://ui.adsabs.harvard.edu/abs/2015ApJ...807..143C} {807, 143}

\bibitem[\protect\citeauthoryear{{Cho} \& {Chae}}{{Cho} \&
  {Chae}}{2020}]{cho20}
{Cho} K.,  {Chae} J.,  2020, \mn@doi [\apjl] {10.3847/2041-8213/ab8295}, \href
  {https://ui.adsabs.harvard.edu/abs/2020ApJ...892L..31C} {892, L31}

\bibitem[\protect\citeauthoryear{{Christensen-Dalsgaard}}{{Christensen-Dalsgaard}}{2002}]{chr02}
{Christensen-Dalsgaard} J.,  2002, \mn@doi [Reviews of Modern Physics]
  {10.1103/RevModPhys.74.1073}, \href
  {https://ui.adsabs.harvard.edu/abs/2002RvMP...74.1073C} {74, 1073}

\bibitem[\protect\citeauthoryear{{Cooper}, {Nakariakov}  \&
  {Tsiklauri}}{{Cooper} et~al.}{2003}]{coo03}
{Cooper} F.~C.,  {Nakariakov} V.~M.,   {Tsiklauri} D.,  2003, \mn@doi [\aap]
  {10.1051/0004-6361:20021556}, \href
  {https://ui.adsabs.harvard.edu/abs/2003A&A...397..765C} {397, 765}

\bibitem[\protect\citeauthoryear{{Culhane} et~al.,}{{Culhane}
  et~al.}{2007}]{cul07}
{Culhane} J.~L.,  et~al., 2007, \mn@doi [\solphys] {10.1007/s01007-007-0293-1},
  \href {https://ui.adsabs.harvard.edu/abs/2007SoPh..243...19C} {243, 19}

\bibitem[\protect\citeauthoryear{{De Moortel}}{{De Moortel}}{2009}]{dem09}
{De Moortel} I.,  2009, \mn@doi [\ssr] {10.1007/s11214-009-9526-5}, \href
  {https://ui.adsabs.harvard.edu/abs/2009SSRv..149...65D} {149, 65}

\bibitem[\protect\citeauthoryear{{De Moortel} \& {Hood}}{{De Moortel} \&
  {Hood}}{2003}]{dem03}
{De Moortel} I.,  {Hood} A.~W.,  2003, \mn@doi [\aap]
  {10.1051/0004-6361:20030984}, \href
  {https://ui.adsabs.harvard.edu/abs/2003A&A...408..755D} {408, 755}

\bibitem[\protect\citeauthoryear{{De Moortel} \& {Hood}}{{De Moortel} \&
  {Hood}}{2004}]{dem04}
{De Moortel} I.,  {Hood} A.~W.,  2004, \mn@doi [\aap]
  {10.1051/0004-6361:20034233}, \href
  {https://ui.adsabs.harvard.edu/abs/2004A&A...415..705D} {415, 705}

\bibitem[\protect\citeauthoryear{{De Moortel} \& {Nakariakov}}{{De Moortel} \&
  {Nakariakov}}{2012}]{dem12}
{De Moortel} I.,  {Nakariakov} V.~M.,  2012, \mn@doi [Philosophical
  Transactions of the Royal Society of London Series A]
  {10.1098/rsta.2011.0640}, \href
  {https://ui.adsabs.harvard.edu/abs/2012RSPTA.370.3193D} {370, 3193}

\bibitem[\protect\citeauthoryear{{De Moortel}, {Ireland}, {Hood}  \&
  {Walsh}}{{De Moortel} et~al.}{2002}]{dem02}
{De Moortel} I.,  {Ireland} J.,  {Hood} A.~W.,   {Walsh} R.~W.,  2002, \mn@doi
  [\aap] {10.1051/0004-6361:20020436}, \href
  {https://ui.adsabs.harvard.edu/abs/2002A&A...387L..13D} {387, L13}

\bibitem[\protect\citeauthoryear{{De Pontieu} \& {McIntosh}}{{De Pontieu} \&
  {McIntosh}}{2010}]{dep10}
{De Pontieu} B.,  {McIntosh} S.~W.,  2010, \mn@doi [\apj]
  {10.1088/0004-637X/722/2/1013}, \href
  {https://ui.adsabs.harvard.edu/abs/2010ApJ...722.1013D} {722, 1013}

\bibitem[\protect\citeauthoryear{{De Pontieu}, {Erd{\'e}lyi}  \& {De
  Moortel}}{{De Pontieu} et~al.}{2005}]{dep05}
{De Pontieu} B.,  {Erd{\'e}lyi} R.,   {De Moortel} I.,  2005, \mn@doi [\apjl]
  {10.1086/430345}, \href
  {https://ui.adsabs.harvard.edu/abs/2005ApJ...624L..61D} {624, L61}

\bibitem[\protect\citeauthoryear{{Del Zanna} \& {Mason}}{{Del Zanna} \&
  {Mason}}{2003}]{del03}
{Del Zanna} G.,  {Mason} H.~E.,  2003, \mn@doi [\aap]
  {10.1051/0004-6361:20030791}, \href
  {https://ui.adsabs.harvard.edu/abs/2003A&A...406.1089D} {406, 1089}

\bibitem[\protect\citeauthoryear{{Edwin} \& {Roberts}}{{Edwin} \&
  {Roberts}}{1982}]{edw82}
{Edwin} P.~M.,  {Roberts} B.,  1982, \mn@doi [\solphys] {10.1007/BF00170986},
  \href {https://ui.adsabs.harvard.edu/abs/1982SoPh...76..239E} {76, 239}

\bibitem[\protect\citeauthoryear{{Edwin} \& {Roberts}}{{Edwin} \&
  {Roberts}}{1983}]{edw83}
{Edwin} P.~M.,  {Roberts} B.,  1983, \mn@doi [\solphys] {10.1007/BF00196186},
  \href {https://ui.adsabs.harvard.edu/abs/1983SoPh...88..179E} {88, 179}

\bibitem[\protect\citeauthoryear{{Handy} et~al.,}{{Handy} et~al.}{1999}]{han99}
{Handy} B.~N.,  et~al., 1999, \mn@doi [\solphys] {10.1023/A:1005166902804},
  \href {https://ui.adsabs.harvard.edu/abs/1999SoPh..187..229H} {187, 229}

\bibitem[\protect\citeauthoryear{{Jess}, {De Moortel}, {Mathioudakis},
  {Christian}, {Reardon}, {Keys}  \& {Keenan}}{{Jess} et~al.}{2012}]{jes12}
{Jess} D.~B.,  {De Moortel} I.,  {Mathioudakis} M.,  {Christian} D.~J.,
  {Reardon} K.~P.,  {Keys} P.~H.,   {Keenan} F.~P.,  2012, \mn@doi [\apj]
  {10.1088/0004-637X/757/2/160}, \href
  {https://ui.adsabs.harvard.edu/abs/2012ApJ...757..160J} {757, 160}

\bibitem[\protect\citeauthoryear{{Kiddie}, {De Moortel}, {Del Zanna},
  {McIntosh}  \& {Whittaker}}{{Kiddie} et~al.}{2012}]{kid12}
{Kiddie} G.,  {De Moortel} I.,  {Del Zanna} G.,  {McIntosh} S.~W.,
  {Whittaker} I.,  2012, \mn@doi [\solphys] {10.1007/s11207-012-0042-5}, \href
  {https://ui.adsabs.harvard.edu/abs/2012SoPh..279..427K} {279, 427}

\bibitem[\protect\citeauthoryear{{Klimchuk}, {Tanner}  \& {De
  Moortel}}{{Klimchuk} et~al.}{2004}]{klim04}
{Klimchuk} J.~A.,  {Tanner} S.~E.~M.,   {De Moortel} I.,  2004, \mn@doi [\apj]
  {10.1086/425122}, \href
  {https://ui.adsabs.harvard.edu/abs/2004ApJ...616.1232K} {616, 1232}

\bibitem[\protect\citeauthoryear{{Kolotkov} \& {Nakariakov}}{{Kolotkov} \&
  {Nakariakov}}{2022}]{kol22}
{Kolotkov} D.~Y.,  {Nakariakov} V.~M.,  2022, \mn@doi [\mnras]
  {10.1093/mnrasl/slac054}, \href
  {https://ui.adsabs.harvard.edu/abs/2022MNRAS.514L..51K} {514, L51}

\bibitem[\protect\citeauthoryear{{Krishna Prasad}, {Banerjee}  \& {Van
  Doorsselaere}}{{Krishna Prasad} et~al.}{2014}]{kri14}
{Krishna Prasad} S.,  {Banerjee} D.,   {Van Doorsselaere} T.,  2014, \mn@doi
  [\apj] {10.1088/0004-637X/789/2/118}, \href
  {https://ui.adsabs.harvard.edu/abs/2014ApJ...789..118K} {789, 118}

\bibitem[\protect\citeauthoryear{{Krishna Prasad}, {Jess}  \&
  {Khomenko}}{{Krishna Prasad} et~al.}{2015}]{kri15}
{Krishna Prasad} S.,  {Jess} D.~B.,   {Khomenko} E.,  2015, \mn@doi [\apjl]
  {10.1088/2041-8205/812/1/L15}, \href
  {https://ui.adsabs.harvard.edu/abs/2015ApJ...812L..15K} {812, L15}

\bibitem[\protect\citeauthoryear{{Krishna Prasad}, {Jess}, {Klimchuk}  \&
  {Banerjee}}{{Krishna Prasad} et~al.}{2017}]{kri17}
{Krishna Prasad} S.,  {Jess} D.~B.,  {Klimchuk} J.~A.,   {Banerjee} D.,  2017,
  \mn@doi [\apj] {10.3847/1538-4357/834/2/103}, \href
  {https://ui.adsabs.harvard.edu/abs/2017ApJ...834..103K} {834, 103}

\bibitem[\protect\citeauthoryear{{Landi} \& {Feldman}}{{Landi} \&
  {Feldman}}{2008}]{lan08}
{Landi} E.,  {Feldman} U.,  2008, \mn@doi [\apj] {10.1086/523629}, \href
  {https://ui.adsabs.harvard.edu/abs/2008ApJ...672..674L} {672, 674}

\bibitem[\protect\citeauthoryear{{Landi} \& {Klimchuk}}{{Landi} \&
  {Klimchuk}}{2010}]{lan10}
{Landi} E.,  {Klimchuk} J.~A.,  2010, \mn@doi [\apj]
  {10.1088/0004-637X/723/1/320}, \href
  {https://ui.adsabs.harvard.edu/abs/2010ApJ...723..320L} {723, 320}

\bibitem[\protect\citeauthoryear{{Lemen} et~al.,}{{Lemen} et~al.}{2012}]{lem12}
{Lemen} J.~R.,  et~al., 2012, \mn@doi [\solphys] {10.1007/s11207-011-9776-8},
  \href {https://ui.adsabs.harvard.edu/abs/2012SoPh..275...17L} {275, 17}

\bibitem[\protect\citeauthoryear{{McIntosh}, {de Pontieu}, {Carlsson},
  {Hansteen}, {Boerner}  \& {Goossens}}{{McIntosh} et~al.}{2011}]{mci11}
{McIntosh} S.~W.,  {de Pontieu} B.,  {Carlsson} M.,  {Hansteen} V.,  {Boerner}
  P.,   {Goossens} M.,  2011, \mn@doi [\nat] {10.1038/nature10235}, \href
  {https://ui.adsabs.harvard.edu/abs/2011Natur.475..477M} {475, 477}

\bibitem[\protect\citeauthoryear{{Nakariakov}, {Verwichte}, {Berghmans}  \&
  {Robbrecht}}{{Nakariakov} et~al.}{2000}]{nak00}
{Nakariakov} V.~M.,  {Verwichte} E.,  {Berghmans} D.,   {Robbrecht} E.,  2000,
  \aap, \href {https://ui.adsabs.harvard.edu/abs/2000A&A...362.1151N} {362,
  1151}

\bibitem[\protect\citeauthoryear{{Ofman}, {Nakariakov}  \& {Sehgal}}{{Ofman}
  et~al.}{2000}]{ofman00}
{Ofman} L.,  {Nakariakov} V.~M.,   {Sehgal} N.,  2000, \mn@doi [\apj]
  {10.1086/308691}, \href
  {https://ui.adsabs.harvard.edu/abs/2000ApJ...533.1071O} {533, 1071}

\bibitem[\protect\citeauthoryear{{Schrijver} et~al.,}{{Schrijver}
  et~al.}{1999}]{sch99}
{Schrijver} C.~J.,  et~al., 1999, \mn@doi [\solphys] {10.1023/A:1005194519642},
  \href {https://ui.adsabs.harvard.edu/abs/1999SoPh..187..261S} {187, 261}

\bibitem[\protect\citeauthoryear{{Uritsky}, {Davila}, {Viall}  \&
  {Ofman}}{{Uritsky} et~al.}{2013}]{uri13}
{Uritsky} V.~M.,  {Davila} J.~M.,  {Viall} N.~M.,   {Ofman} L.,  2013, \mn@doi
  [\apj] {10.1088/0004-637X/778/1/26}, \href
  {https://ui.adsabs.harvard.edu/abs/2013ApJ...778...26U} {778, 26}

\bibitem[\protect\citeauthoryear{{Wang}}{{Wang}}{2016}]{wan16}
{Wang} T.~J.,  2016, \mn@doi [Geophysical Monograph Series]
  {10.1002/9781119055006.ch23}, \href
  {https://ui.adsabs.harvard.edu/abs/2016GMS...216..395W} {216, 395}

\bibitem[\protect\citeauthoryear{{Wang}, {Ofman}, {Davila}  \&
  {Mariska}}{{Wang} et~al.}{2009}]{wan09}
{Wang} T.~J.,  {Ofman} L.,  {Davila} J.~M.,   {Mariska} J.~T.,  2009, \mn@doi
  [\aap] {10.1051/0004-6361/200912534}, \href
  {https://ui.adsabs.harvard.edu/abs/2009A&A...503L..25W} {503, L25}

\bibitem[\protect\citeauthoryear{{Wang}, {Ofman}, {Yuan}, {Reale}, {Kolotkov}
  \& {Srivastava}}{{Wang} et~al.}{2021}]{wan21}
{Wang} T.,  {Ofman} L.,  {Yuan} D.,  {Reale} F.,  {Kolotkov} D.~Y.,
  {Srivastava} A.~K.,  2021, \mn@doi [\ssr] {10.1007/s11214-021-00811-0}, \href
  {https://ui.adsabs.harvard.edu/abs/2021SSRv..217...34W} {217, 34}

\bibitem[\protect\citeauthoryear{{Young}}{{Young}}{2023}]{you23}
{Young} P.~R.,  2023, \mn@doi [\apj] {10.3847/1538-4357/ad0548}, \href
  {https://ui.adsabs.harvard.edu/abs/2023ApJ...958...40Y} {958, 40}

\bibitem[\protect\citeauthoryear{{Young} \& {Viall}}{{Young} \&
  {Viall}}{2022}]{you22}
{Young} P.~R.,  {Viall} N.~M.,  2022, \mn@doi [\apj]
  {10.3847/1538-4357/ac8472}, \href
  {https://ui.adsabs.harvard.edu/abs/2022ApJ...938...27Y} {938, 27}

\bibitem[\protect\citeauthoryear{{Young} et~al.,}{{Young} et~al.}{2007}]{you07}
{Young} P.~R.,  et~al., 2007, \mn@doi [\pasj] {10.1093/pasj/59.sp3.S857}, \href
  {https://ui.adsabs.harvard.edu/abs/2007PASJ...59S.857Y} {59, S857}

\bibitem[\protect\citeauthoryear{{Yuan}, {Sych}, {Reznikova}  \&
  {Nakariakov}}{{Yuan} et~al.}{2014}]{yua14}
{Yuan} D.,  {Sych} R.,  {Reznikova} V.~E.,   {Nakariakov} V.~M.,  2014, \mn@doi
  [\aap] {10.1051/0004-6361/201220208}, \href
  {https://ui.adsabs.harvard.edu/abs/2014A&A...561A..19Y} {561, A19}

\bibitem[\protect\citeauthoryear{{Zavershinskii}, {Kolotkov}, {Nakariakov},
  {Molevich}  \& {Ryashchikov}}{{Zavershinskii} et~al.}{2019}]{zav19}
{Zavershinskii} D.~I.,  {Kolotkov} D.~Y.,  {Nakariakov} V.~M.,  {Molevich}
  N.~E.,   {Ryashchikov} D.~S.,  2019, \mn@doi [Physics of Plasmas]
  {10.1063/1.5115224}, \href
  {https://ui.adsabs.harvard.edu/abs/2019PhPl...26h2113Z} {26, 082113}

\bibitem[\protect\citeauthoryear{{Zhao}, {Chen}, {Hartlep}  \&
  {Kosovichev}}{{Zhao} et~al.}{2015}]{zha15}
{Zhao} J.,  {Chen} R.,  {Hartlep} T.,   {Kosovichev} A.~G.,  2015, \mn@doi
  [\apjl] {10.1088/2041-8205/809/1/L15}, \href
  {https://ui.adsabs.harvard.edu/abs/2015ApJ...809L..15Z} {809, L15}

\bibitem[\protect\citeauthoryear{{Zhao}, {Felipe}, {Chen}  \&
  {Khomenko}}{{Zhao} et~al.}{2016}]{zha16}
{Zhao} J.,  {Felipe} T.,  {Chen} R.,   {Khomenko} E.,  2016, \mn@doi [\apjl]
  {10.3847/2041-8205/830/1/L17}, \href
  {https://ui.adsabs.harvard.edu/abs/2016ApJ...830L..17Z} {830, L17}

\makeatother
\end{thebibliography}

\end{document}